\documentstyle[12pt]{article}

\def\be{\begin{equation}}
\def\ee{\end{equation}}
\def\bea{\begin{eqnarray}}
\def\eea{\end{eqnarray}}
\def\b{\bar}

\def\l{\lambda}

\def\m{\mu}
\def\n{\nu}

\def\t{\tau}

\def\~{\tilde}

\def\z{\zeta}
\def\Z{{\b\zeta}}
\def\Y{{\b Y}}

\def\`{\dot}
\begin{document}
\title{Twisting Lightlike Solutions of the Kerr-Schild Class}
\author{Alexander Burinskii \\
Russian Academy of Sciences, Moscow
\and
Giulio Magli \\
Dipartimento di Matematica del Politecnico di Milano, Italy}
\date{gr-qc/0008055, August 2000}
\maketitle
\begin{abstract}
Using a complex representation of the Debney--Kerr--Schild (DKS)
solutions and the Kerr theorem we give a method
to construct boosted Kerr geometries.
In the ultrarelativistic case this method yelds
twisting solutions having,
contrary to the known pp-wave limiting solutions,
a non-zero value of the total angular momentum.
The solutions show that twist plays a crucial role in
removing singularity and smoothing shock
wave in the ultrarelativistic limit.
Two different physical situations are discussed.
\end{abstract}
The problem of boosting of the black hole solutions received considerable
attention in connection with different problems, like
interaction between black holes and estimation of the gravitational
interparticle interaction at very high energies.
\par
First results in this field were obtained by Aichelburg and Sexl [1], who
considered the boosting of the Schwarzschild solution.
Due to the singular character of Lorentz transformations at $v=c$,
many difficulties appear when the ultrarelativistic limit is involved.
This singularity can,
{\it a priori}, lead to different limiting results depending on the
performed limiting procedure. Nevertheless, investigations of many authors
led to a general conclusion that black hole solutions
turn, in the ultrarelativistic limits,
into singular pp-waves (for references see [5]).
\par
In the case of the boosting of the rotating Kerr BH, the pp-wave
ultrarelativistic solution looses one of the main properties
of the original solution, namely twist of
the principal null congruence.
In the same time, it is known that the Kerr-Schild class contains
twisting lightlike solutions. The aim of this work is to show that there
exist twisting lightlike solutions which can be considered as
 ultrarelativistic limits of the Kerr BH.
\par
The approach used here is based on the DKS formalism [2] and on
the Kerr theorem.
It gives a possibility of obtaining exact and explicit expressions
for the boosted Kerr geometry by arbitrary values and orientations of the
boost with respect to the angular momentum.
\par
The general Kerr-Schild metric is
$g_{\m\n} =\eta_{\m\n} + 2 h e^3_{\m} e^3_{\n} $,
where $h$ is a scalar function and $ e^3 $
the principal null direction, given
in null coordinates $(\z ,\Z , u, v)$ by
$e^3 = du+ \Y d \z  + Y d \Z - Y \Y d v $ ,
where $Y(x)$ is a complex function.
The Kerr theorem gives a rule to construct all geodesic,
shear free congruences:
the general geodesic, shear-free null congruence in
Minkowski space is defined by a function $Y$ which is a solution of the
equation
\begin{equation}
F (\l_1,\l_2 ,Y) = 0 \ ,
\end{equation}
where  $F$ is an arbitrary analytic
function of the {\it projective twistor coordinates}
$\l_1 = \z - Y v, \qquad \l_2 =u + Y \Z, \qquad Y$.
The quantity $\tilde r := - d F / d Y $
is a complex radial distance, and
singularities of the metric  can be defined as the
caustics of the congruence given by the system of equations
\begin{equation}
F=0, \qquad
d F / d Y =0 \ .
\end{equation}
The Kerr solution belongs to the sub--class of
solutions having singularities
contained in a bounded region.
In this case the function $F$ must be at most quadratic in $Y$.
The solutions of the equations (8)
can be found in this case in explicit form, and
correspond to the Kerr solution up to a Lorentz boost and a
shift of the origin.
Newman [3] constructed a complex representation
which allows to represent the Kerr solution as a
retarded-time field,
generated by a complex source propagating along a complex world-line.
In the complex version of the Kerr theorem [4]
the function $F$ depends on
the coordinates of the complex
world-line $ x_0 (\t)= (\z_0, \Z_0, u_0 ,v_0) \in CM^4$.
parameterized by a complex time parameter $\t=t+i\sigma$.
\par
The one-to-one correspondence between
straight lines in complex Minkowski space and
the class of the DKS solutions having singularities contained in a
bounded region allows us to boost the Kerr solution
{\it via} the DKS formalism.
The general case of a solution with a boost corresponds to
a straight, complex
world line with 3-velocity $\vec V$ in $CM^4$
\begin{equation}
x_0^\m (\t) = x_0^\m (0) + \xi^\m \t; \qquad \xi^\m = (1,\vec V)\ .
\end{equation}
It allows one to perform DKS-machinery and obtain solutions corresponding to
parameters of the complex world line.
On this way all the versions of the boosting of the Kerr BH can be considered
in explicit form.
\par
In particular, the most interesting case is that a boost collinear to
the angular momentum. In this case,
the Kerr singular ring grows as $a/\sqrt{1-v^2}$ and tends to infinity
in the ultrarelativistic limit. In this limit,
the function $F$ acquires the form
$F=x+iy - Y (z-ia -t)$. It follows that
the complex radial distance is
\begin{equation}
\tilde r = - dF/dY = z-ia-t ,
\end{equation}
and therefore the metric has no singularity if $a$ is non-zero.
Two different physical situations must be considered in this case:

i) The field of a lightlike particle with a non-zero helicity $J=m_0 a_0=ma$.
The rest mass $m_0$ must be put equal to zero in the limit, so that
$a_0 \to  \infty$. The relativistic parameters $m$ and $a$ are kept constant,
$h=m\frac{z-t}{(z-t)^2+a^2}$, so that the singularity on the front is absent.

ii) Relativistic boost of a particle with a finite rest mass $m_0$.
In this case $m= m_0/\sqrt {1-v^2}$ grows under the effect of the boost,
and consequently, $a= J/m = (J/m_0)\sqrt{1-v^2}$
is going to zero. It follows that $h$
grows with the boost forming a wave with amplitude
of order $\frac {m_0}{a_0(1-v^2)}$.
However, $m_0/a_0 $ is of order $10^{-44}$ for an electron, so that
this effect can be observed only at
$1-v^2 \sim m_0/a_0 \sim 10^{-44}$.
\par
In conclusion, in this approach twist plays a crucial role in
removing singularity and smoothing shock
wave in the ultrarelativistic limit.
\par

\end{document}